\documentclass[10pt]{article}
\newcommand{\keywords}[1]{\textbf{Keywords:} #1}

\usepackage[margin=1in]{geometry}

\usepackage[T1]{fontenc}
\usepackage{lmodern}
\usepackage{microtype}
\usepackage{authblk}
\usepackage{amsmath,amssymb,amsthm}
\usepackage{booktabs,multirow,siunitx}
\usepackage[table]{xcolor}
\usepackage{graphicx}
\usepackage{tikz}
\usepackage{listings}
\usetikzlibrary{positioning,fit,arrows.meta,calc,decorations.pathreplacing}

\usepackage[numbers]{natbib}
\usepackage[hidelinks]{hyperref}
\usepackage{authblk}

\input{./definitions}

\begin{document}


\title{Large-Scale Regularized Matching on GPU Clusters}

\author[1]{Aida Rahmattalabi\thanks{Three authors contributed equally to this research.}}
\author[2]{Gregory Dexter$^\ast$\thanks{Work completed at Linkedin}}
\author[1]{Sanjana Garg$^\ast$}
\author[3]{Qinquan Song$^\dagger$}
\author[1]{Shenyinying Tu}
\author[1]{Yuan Gao}
\author[1]{Zhipeng Wang}
\author[1]{Rahul Mazumder}

\affil[1]{LinkedIn Corporation, USA}
\affil[2]{Nubank, USA}
\affil[3]{OpenAI, USA}

\date{}
\maketitle

\begin{abstract}
Production decision systems such as ad allocation and content matching involve millions of users and thousands of items, reducing to large-scale linear programs with a common structure: sparse constraint matrices with diagonal block structure across sources (users), solved repeatedly on recurring cadences over slowly evolving inputs. Three systems gaps stand out. Scale: production instances routinely exceed the memory capacity of GPU solvers such as cuPDLP and D-PDLP under fixed hardware budgets. Temporal instability: solution variability across consecutive runs induces downstream churn and complicates SLAs, a failure mode for which existing solvers provide no explicit control. Extensibility: CPU-based solvers such as DuaLip-Scala converge slowly and couple problem formulation to fixed schemas, making new constraint families expensive to incorporate.

We present a distributed multi-GPU LP solver built natively in PyTorch with systems–algorithm co-design for this structure. It adopts a column-sharded execution model with fused Triton kernels and batched operations to reduce per-iteration overhead. As the number of users grows, only local computation increases, while communication is limited to a single reduction of item-level dual variables to rank 0, yielding near-linear scaling with GPU count at fixed item size. Second, we adopt ridge-regularized LPs from prior work to improve solver stability, introducing a tunable $\gamma$ that provably bounds run-to-run solution drift. This control is not exposed in existing GPU LP solvers. A continuation schedule over $\gamma$ further balances convergence speed and solution fidelity. Finally, we introduce an operator-centric programming model that replaces DuaLip-Scala’s schema-bound interface with three composable primitives (ObjectiveFunction, ProjectionMap, Maximizer), enabling new formulations through local changes without modifying the solve loop or distributed infrastructure. On synthetic matching workloads representative of production usage, our system achieves order-of-magnitude wall-clock speedup over DuaLip-Scala, near-linear multi-GPU scaling (3.86$\times$ on 4 GPUs), while scaling to problem sizes beyond the reach of GPU-based solvers.
\end{abstract}



 \keywords{linear programming, GPU computing, distributed optimization, dual ascent, matching, ridge regularization}



\section{Introduction}

Large-scale linear programs (LPs) underpin production decision systems, where instances with billions of decision variables and hundreds of millions of constraints are solved repeatedly under tight latency and stability requirements. Classical LP solvers such as simplex or interior-point methods do not scale well to this regime: both are bottlenecked by sparse matrix factorizations, which are memory-intensive and difficult to parallelize efficiently at scale. This leads to poor utilization of modern parallel hardware and limits their applicability in large-scale production settings. Another challenge is stability. In these settings, small perturbations in the data can lead to large changes in the solution. To address this, regularization terms are often introduced to encourage proximity to a reference solution (e.g., the prior-day solution) and reduce solution drift. However, in the linear setting, this effectively introduces additional constraints into an already large-scale problem, further increasing computational burden.

The LPs arising in these settings share a structure that existing solvers do not fully exploit. Decision variables partition into per-source blocks (one per user), each subject to simple local constraints—such as simplex or box constraints—that decompose independently across blocks. Coupling constraints (e.g., budgets, pacing limits, and frequency caps) interact with each source block in an element-wise manner, yielding a constraint matrix with diagonal block structure across sources. The resulting matrices are highly sparse: nonzeros occur only at eligible source–destination pairs, and within each block the interaction is diagonal. 

Recently, first-order methods have emerged as a scalable approach for large linear programs. These are iterative primal–dual algorithms, notably Primal-Dual Hybrid Gradient (PDHG)~\cite{applegate2021practical}, whose per-iteration cost reduces to sparse matrix–vector multiplication, making them well-suited to GPUs. This has enabled solvers such as \texttt{cuPDLP.jl}~\cite{lu2025cupdlp} and \texttt{cuPDLPx}~\cite{lu2025cupdlpxenhancedgpubasedfirstorder}, the latter improving convergence via restarted Halpern PDHG. However, these methods are strictly single-device, with memory-bound problem sizes and no path to scale-out. \texttt{D-PDLP}~\cite{li2026dpdlpscalingpdlpdistributed} extends PDHG to multiple GPUs via 2D matrix partitioning and NCCL AllReduce, achieving up to $6\times$ speedup on NVLink-connected H100s, but remains confined to a single NVLink domain and requires synchronous communication at every iteration, with no mechanism for stability control across recurring solves. Crucially, these methods ignore structural properties of production LPs, where variables decompose into independent per-source blocks with simple local constraints, and coupling constraints act element-wise, inducing a diagonal block structure. Existing solvers treat the system as unstructured, missing this decomposition. A complementary line of work, including ECLIPSE~\citep{basu2020eclipse} and DuaLip~\citep{ramanath2021efficient}, exploits this structure via ridge-regularized dual ascent and scales to CPU clusters, but remains limited to fixed schemas and CPU-centric designs, making extensions costly and limiting accelerator efficiency.

We build on DuaLip~\citep{ramanath2021efficient} and  present a distributed multi-GPU LP solver built natively in PyTorch that co-designs algorithm and system components around the diagonal block structure of production matching LPs. Our contributions are:

\begin{enumerate}

\item \textbf{Column-sharded SpMV and projection via NCCL.}
We partition the constraint matrix across GPUs by source columns. Because constraints decompose per source, each GPU computes local primal updates and gradient contributions independently, with no cross-device dependencies during computation. Each iteration requires only a reduction and broadcast of destination-level dual variables to and from rank 0, whose size is independent of the number of sources and the column partitioning. As a result, scaling the number of sources increases only local computation while communication remains fixed, enabling near-linear scaling on instances that exceed single-device memory capacity.

\item \textbf{Tunable regularization with stability control.}
We adopt a ridge-regularized formulation from prior work and expose the parameter $\gamma$ as a first-class control, which bounds run-to-run primal drift and directly addresses stability requirements in recurring production solves. A continuation schedule initializes with larger $\gamma$ for fast, stable convergence and anneals it toward the target fidelity.

\item \textbf{Operator-centric programming model.}
We replace DuaLip-Scala’s rigid schemas with three composable primitives: \texttt{ObjectiveFunction}, \texttt{Maximizer} and \texttt{ProjectionMap}. New LP formulations require only local implementation of the objective and gradient, while the solve loop and distributed infrastructure remain unchanged, improving extensibility without affecting performance.
\end{enumerate}

On synthetic matching workloads calibrated to production-scale sizes, our system achieves a $9.1\times$ wall-clock speedup over DuaLip-Scala on a single GPU and exhibits near-linear scaling that improves with the number of sources.

\section{Related Work}

\textit{Large-scale matching and allocation.}
A substantial literature addresses weighted bipartite matching and network-flow problems at scale, with specialized solvers that exploit degree and capacity constraints~\citep{vazirani2001approximation, fleischer2006tight, tsiakis2008optimal}. However, these methods do not naturally support heterogeneous linear side constraints such as pacing, fairness, and frequency caps that arise in production allocation systems~\citep{lobo2024fair, buchbinder2014frequency}. The formulation inherited from ECLIPSE preserves general LP expressivity while exploiting matching structure in the dual, enabling new constraint families to be added without modifying the core solver. Production applications include notification and email allocation~\citep{basu2020eclipse}, ad allocation matching~\citep{naor2018near}, opportunity assignment~\citep{ramanath2021efficient}, and inactive-member promotion on social networks~\citep{acharya2023promoting}.

\textit{First-order methods.}
PDLP~\citep{applegate2021practical} established convergence guarantees for restarted PDHG on general LPs and underpins both \textsc{cuPDLP} variants~\citep{lu2024cupdlpjlgpuimplementationrestarted, lu2025cupdlpxenhancedgpubasedfirstorder}. These methods reduce per-iteration computation to generic sparse matrix–vector products but do not exploit problem structure. In addition, their analysis does not address solution stability across repeated solves, a key requirement in production decision systems.

\textit{Ridge-regularized dual ascent.}
\citet{basu2020eclipse} showed that $\ell_2$ regularization induces smoothness in the dual and enables scalable optimization for extreme-scale allocation, identifying the diagonal-block structure of web-scale matching LPs as the key property for distributed decomposition. \citet{ramanath2021efficient} operationalized these ideas in a Scala/Spark system (\textsc{DuaLip}), with efficient projections for simplex and box constraints. However, the system is CPU-centric and does not support accelerator-friendly sparse layouts or fine-grained collective communication. It also exposes only two fixed schemas, making new constraint families require non-local code changes.

\textit{GPU-based LP solvers.}
\citet{lu2024cupdlpjlgpuimplementationrestarted} showed that restarted PDHG can match commercial solvers when instances fit in GPU memory, but is fundamentally limited by a single-device memory ceiling. \citet{lu2025cupdlpxenhancedgpubasedfirstorder} improved convergence via restarted Halpern PDHG, achieving $2.5\times$--$5\times$ speedups over \textsc{cuPDLP}, while retaining the same memory constraint. \citet{li2026dpdlpscalingpdlpdistributed} extended PDHG to multiple GPUs via 2D matrix partitioning and NCCL-based synchronization, achieving up to $6\times$ speedup on eight NVLink-connected H100s within a single server. However, these methods still fail to exploit the structure of production matching LPs. First, all reported experiments are confined to a single NVLink domain,  leaving multi-node scalability unaddressed. Second, treating the constraint matrix as unstructured forces a generic 2D partitioning that requires two synchronous AllReduces per iteration along the row and column grid. In our setting, the column/source decomposition makes the per-iteration reduce volume scale with the constraint dimension (destinations) and independent of the decision-variable dimension (sources) which is the larger of the two in production matching workloads.

\textit{High-performance tensor frameworks.}
Our implementation builds on PyTorch~\citep{paszke2019pytorch}, leveraging sparse-dense and fused kernels, batched dense operations, and \texttt{torch.distributed} with an NCCL backend. This provides a flexible execution substrate for structured sparse linear algebra, bridging the gap between LP-specific algorithm design and general-purpose GPU programming. Further, it enables a unified implementation in which the solve loop and projection operators are shared across formulations while remaining efficient on distributed GPU hardware.

\section{Problem Setting}
\label{sec:problem}
This section formalizes the ridge-regularized LP used throughout the paper, derives its dual, and specializes both to the matching workloads that motivate our system. The resulting computational object consists of two sparse matrix–vector products and a blockwise projection over a constraint matrix with diagonal block structure, which forms the basis for the design in Section~\ref{sec:gpu}.

\subsection{Ridge-Regularized LP}
\label{sec:preliminaries}
We consider the regularized LP problem
\begin{equation}
\label{eq:primal}
\min_{\mathbf{x} \in \mathcal{C}}\quad
   \mathbf{c}^{\top}\mathbf{x}
   + \frac{\gamma}{2}\,\|\mathbf{x}\|_{2}^{2}
\quad\text{s.t.}\quad
   \mathbf{A}\mathbf{x} \le \mathbf{b},
\end{equation}
with primal variable $\mathbf{x}\in\mathbb{R}^{n}$, sparse coupling
matrix $\mathbf{A}\in\mathbb{R}^{m\times n}$, right-hand side
$\mathbf{b}\in\mathbb{R}^{m}$, and regularization parameter
$\gamma>0$. The set $\mathcal{C}\subseteq\mathbb{R}^{n}$ encodes
\emph{simple constraints} that decompose blockwise across sources and admits an efficient projection oracle $\Pi_{\mathcal{C}}$; the inequality $\mathbf{A}\mathbf{x}\le\mathbf{b}$ encodes the \emph{coupling constraints} that bind blocks together. This distinction is operationally central: simple constraints are absorbed into the primal subproblem via projection and do not appear in the dual, while coupling constraints are the only ones associated with explicit dual variables. The ridge term induces strong convexity in the primal, yielding a smooth dual with a fixed step size that can be derived analytically from $\mathbf{A}$ and $\gamma$
and it preserves the blockwise separability of the inner minimization, which the column-sharded execution layer of Section~\ref{sec:gpu} relies on.

\subsection{Dual Formulation}
\label{sec:dual}

Let $\boldsymbol{\lambda}\in\mathbb{R}^{m}_{\ge 0}$ denote the dual
variables associated with the constraint $\mathbf{A}\mathbf{x}\le\mathbf{b}$. The Lagrangian dual is given by
\begin{equation}
\label{eq:dual}
g(\boldsymbol{\lambda})
= \min_{\mathbf{x}\in\mathcal{C}}
   \left\{
      \mathbf{c}^{\top}\mathbf{x}
      + \frac{\gamma}{2}\|\mathbf{x}\|_{2}^{2}
      + \boldsymbol{\lambda}^{\top}(\mathbf{A}\mathbf{x}-\mathbf{b})
   \right\}.
\end{equation}
Strong convexity of the primal objective (due to $\gamma>0$) implies strong duality; in other words, solving the dual recovers the primal optimum with no duality gap. The inner minimization decomposes over the blocks of $\mathcal{C}$ and admits a closed-form expression as a projection of an unconstrained quadratic minimizer:
\begin{equation}
\label{eq:primal-recovery}
\mathbf{x}^{*}_{\gamma}(\boldsymbol{\lambda})
= \Pi_{\mathcal{C}}\!\left(
     -\tfrac{1}{\gamma}\bigl(\mathbf{A}^{\top}\boldsymbol{\lambda} + \mathbf{c}\bigr)
   \right).
\end{equation}

By Danskin’s theorem, the dual is differentiable with gradient
\begin{equation}
\label{eq:dual-gradient}
\nabla g(\boldsymbol{\lambda})
= \mathbf{A}\mathbf{x}^{*}_{\gamma}(\boldsymbol{\lambda}) - \mathbf{b}.
\end{equation}

Equations~\eqref{eq:primal-recovery}--\eqref{eq:dual-gradient}
isolate the three dominant per-iteration operations: the sparse
matrix–vector products $\mathbf{A}^{\top}\boldsymbol{\lambda}$ and
$\mathbf{A}\mathbf{x}^{*}_{\gamma}$, and the projection
$\Pi_{\mathcal{C}}$. Their cost is fully determined by the structure
of $\mathbf{A}$, which we exploit in the next section.

\subsection{Matching Workloads}
\label{sec:matching}

The LPs arising in our target workloads share a structure that the rest of the paper exploits. Let $x_{ij}$ denote the assignment of source $i \in [I]$ (a user) to destination $j \in [J]$ (a campaign, item, or slot), and let $E \subseteq [I] \times [J]$ denote the set of eligible pairs. We define $x_{ij} = 0$ for $(i,j) \notin E$ and stack the per-source blocks $\mathbf{x}_i = (x_{i1}, \ldots, x_{iJ})^{\top}$ into vector $\mathbf{x} = (\mathbf{x}_1; \ldots; \mathbf{x}_I)$.

\textit{Simple constraints.}
Each source has a local constraint, typically a unit simplex constraint $\sum_j x_{ij} \le 1$ with $x_{ij} \ge 0$ for $(i,j)\in E$, or a box variant, which decomposes independently across sources. These correspond to the simple constraints $\mathbf{x}\in\mathcal{C}$ in~\eqref{eq:primal}, where each block $\mathbf{x}_i$ lies in a separate polytope $\mathcal{C}_i$ and the projection operator $\Pi_{\mathcal{C}}$ decomposes as a Cartesian product of per-block projections~\cite{ramanath2021efficient}. Crucially, these constraints do not introduce dual variables; only the coupling constraints do. As a result, the number of explicit dual variables scales as $O(J)$ rather than $O(I+J)$, with $J \ll I$ in production settings, which directly determines the communication volume in Section~\ref{sec:gpu}.

\textit{Coupling constraints.}
Cross-source requirements, including destination-level budgets, pacing, frequency caps, and fairness limits, couple all sources eligible for a given destination. A canonical example is destination capacity,
\begin{equation}
\label{eq:dest-capacity}
\sum_{i\in[I]} a_{ij}\, x_{ij} \le d_j
\quad\forall j\in[J],
\end{equation}
which interacts with each source block in an element-wise manner. With $\mathbf{x}$ stacked in source-major order, this yields a coupling matrix formed by horizontally concatenated diagonal blocks. We allow multiple such constraint families to coexist, producing the structure exploited in Section~\ref{sec:gpu}.
\begin{definition}[Matching coupling matrix]
\label{def:matching-A}
For $m$ coupling-constraint families indexed by $k\in[m]$ and
$\mathbf{x}=(\mathbf{x}_1;\ldots;\mathbf{x}_I)\in\mathbb{R}^{IJ}$,
the coupling matrix $\mathbf{A}\in\mathbb{R}^{mJ\times IJ}$ is
\begin{equation}
\mathbf{A} =
\begin{bmatrix}
\mathbf{D}_{11} & \mathbf{D}_{12} & \cdots & \mathbf{D}_{1I} \\
\mathbf{D}_{21} & \mathbf{D}_{22} & \cdots & \mathbf{D}_{2I} \\
\vdots          & \vdots          & \ddots & \vdots          \\
\mathbf{D}_{m1} & \mathbf{D}_{m2} & \cdots & \mathbf{D}_{mI}
\end{bmatrix},
\end{equation}
where $\mathbf{D}_{ki}\in\mathbb{R}^{J\times J}\text{ diagonal for every }(k,i).$ Each block $\mathbf{D}_{ki}$ acts element-wise on
$\mathbf{x}_i$, and is zero on coordinates $j$ with $(i,j)\notin E$.
\end{definition}
This object generalizes the ECLIPSE formulation~\cite{basu2020eclipse}, which supported a single matching block, to an arbitrary number of constraint families, each sharing a diagonal block-across-sources structure. It is sparse in two ways: \emph{structurally}, since off-diagonal entries within each $\mathbf{D}_{ki}$ are zero by construction, and \emph{structurally in the data}, since $a_{ij}=0$ for ineligible pairs. The rest of the paper treats this dual sparsity as the central design constraint. Section~\ref{sec:gpu} encodes it in a compact CSC layout that materializes only the $J$ diagonal entries of each block and shards $\mathbf{A}$ across GPUs along the source axis, so per-iteration communication scales with the dual dimension $mJ$ rather than the number of nonzeros or sources. 

\section{GPU System Design}
\label{sec:gpu}

The dual ascent of Section~\ref{sec:problem} reduces each iteration to two sparse matrix--vector products and a blockwise projection. This section describes how we implement these primitives on a multi-GPU cluster. Our goal is to approximate the performance of a custom C++/CUDA solver while retaining a lightweight implementation built on PyTorch sparse tensors and \texttt{torch.distributed}, so that the same code path applies across all formulations supported by the programming model of Section~\ref{sec:architecture}.

We make four design choices, each driven by structural properties of the matching LPs in Definition~\ref{def:matching-A}: a sparse tensor layout preserving block-diagonal and within-block sparsity (\S\ref{sec:gpu-layout}); a bucketed batching scheme that maps per-block projections to a small number of high-occupancy kernel launches (\S\ref{sec:gpu-batching}); a fused Triton kernel for simplex projection that collapses a multi-op PyTorch pipeline into a single kernel (\S\ref{sec:gpu-triton}); and a column-sharded execution layer whose per-iteration communication depends only on the number of coupling constraints (\S\ref{sec:gpu-dist}).

\subsection{Sparse Tensor Layout}
\label{sec:gpu-layout}
The constraint matrix of Definition~\ref{def:matching-A} exhibits two sources of sparsity: \emph{structured} sparsity from the diagonal block form across sources, and \emph{unstructured} sparsity within each block, since only a subset of source--destination pairs are eligible. We exploit both by collapsing the diagonal block structure into a compact representation. With a single matching constraint family, $\mathbf{A} = [\mathbf{D}_1, \ldots, \mathbf{D}_I]$, where each $\mathbf{D}_i \in \mathbb{R}^{J \times J}$ is diagonal, so the nonzero pattern is fully captured by the $J$ diagonal entries per block. We store these in a sparse tensor $\mathcal{T}$ in Compressed Sparse Column (CSC) format with columns indexed by source $i$, so that $\mathcal{T}[:, i] = \mathrm{diag}(\mathbf{D}_i)$ and all variables for a given source are stored contiguously in memory. CSC stores exactly the required information at fixed precision: one column pointer, one row index per nonzero, and one value per nonzero, with no representation of structural zeros within blocks. This layout directly supports the rest of the pipeline. The sparse products $\mathbf{A}\mathbf{x}$ and $\mathbf{A}^{\top}\boldsymbol{\lambda}$ use PyTorch optimized sparse-dense kernels without format conversion, and columns of $\mathcal{T}$ align with the destination partitioning of Basu~et~al.~\cite{basu2020eclipse}, making block-local operations column-local. 

\subsection{Batched Projection}
\label{sec:gpu-batching}

The closed-form primal step~\eqref{eq:primal-recovery} requires projecting each per-source slice of the candidate primal onto its feasible set $\mathcal{C}_i$, a unit simplex or box depending on the formulation. A direct GPU implementation encounters two bottlenecks. Per-source kernel launches incur prohibitive dispatch overhead at scale, while a single dense slab representation of size $[s_{\max} \times \mathrm{num\_sources}]$ eliminates launch overhead but introduces substantial zero-padding waste due to heterogeneous slice lengths. We address both issues with a logarithmic bucketing scheme. Slices are grouped by length into ranges $[2^{t-1}, 2^t)$. Within each bucket, we pack slices into a dense slab padded only to the bucket upper bound, apply a single batched projection kernel, and scatter results back. This bounds padding overhead to at most a factor of two within each bucket, while reducing the number of GPU launches to $1 + \lfloor \log_2 s_{\max} \rfloor$. The scheme integrates directly with PyTorch batched dense kernels. Figure~\ref{fig:batching} reports per-iteration runtime and peak memory as a function of source count, comparing bucketing against the single-slab baseline.

\subsection{Fused Simplex Projection}
\label{sec:gpu-triton}
Within each bucket, the dominant projection in our target workloads is onto the simplex
$\{\mathbf{w} \ge 0,\, \sum_i w_i \le z\}$. The standard algorithm of~\citet{duchi2008efficient}---sorting, prefix sums, threshold selection, recovery of the cutoff index $\rho$ and threshold $\theta$, followed by the final subtract-and-clamp projection---maps naturally to high-level tensor frameworks but compiles into multiple GPU kernel launches per invocation. For block sizes $L$ ranging from hundreds to a few thousand, launch overhead and intermediate memory traffic dominate arithmetic, as temporary tensors (sorted values, prefix sums, and masks) are repeatedly materialized in and read back from global memory. We fuse this pipeline into a single Triton kernel~\citep{tillet2025introducing}. Each program instance processes one column: it loads the slice with masked bounds handling, performs the sort and inclusive scan in registers, recovers the cutoff index $\rho$ via a boolean reduction exploiting the monotone prefix structure of the Duchi condition, computes the threshold $\theta$ through a masked reduction, and applies the final subtract-and-clamp projection. This eliminates intermediate materialization and global memory traffic between stages. For the inequality variant, we add an in-kernel early exit that returns the input unchanged when already feasible, using the same numerical tolerance as the reference implementation to preserve numerical equivalence. The kernel runs in fp32, which best matches Triton’s register-resident sort and scan primitives, and supports column lengths up to 8192. Beyond this limit, execution falls back to the multi-launch implementation to avoid register spilling.

The fused kernel serves as the default projection primitive in the inner loop, reducing each primal update to a single kernel launch and eliminating intermediate materialization in global memory. Figure~\ref{fig:triton_kernel} compares per-iteration runtime and peak memory against the PyTorch-eager baseline across workloads ranging from 1M to 50M sources.

\subsection{Distributed Execution}
\label{sec:gpu-dist}

We launch one OS process per GPU via \texttt{torchrun} and form a single \texttt{torch.distributed} process group over NCCL. The same runtime supports both single-node multi-GPU and multi-node deployments through standard rendezvous environment variables. The CSC tensor $\mathcal{T}$ and corresponding entries of $\mathbf{c}$ are partitioned across devices in a balanced column split, while the dual variable $\boldsymbol{\lambda}$ and constraint vector $\mathbf{b}$ are replicated on every rank. Each rank reads the shared instance directly from the network filesystem and materializes only its local column block, so the full matrix is never assembled on a single device and no startup scatter is required. The column partition aligns with the decomposition induced by the simple constraints. For a fixed $\boldsymbol{\lambda}$, each rank computes its local slice of $\mathbf{x}^{*}_{\gamma}(\boldsymbol{\lambda})$ via~\eqref{eq:primal-recovery} and its contribution to $\nabla g(\boldsymbol{\lambda})$ independently, with no cross-device dependencies during computation. Each iteration proceeds as follows. First, each rank computes its local gradient contribution, objective value, and regularization term on-device. Next, a reduce-to-rank-0 (SUM) aggregates the gradient vector of size $|\boldsymbol{\lambda}|$, while scalar reductions aggregate the objective and regularization values. Rank 0 then performs the accelerated gradient update, producing the next iterate and momentum vector, followed by projection onto the nonnegative orthant for inequality coordinates. Finally, the updated dual variables are broadcast from rank 0 to all ranks for the next iteration.

The per-iteration communication volume therefore consists of one $|\boldsymbol{\lambda}|$-sized reduction, one $|\boldsymbol{\lambda}|$-sized broadcast, and $O(1)$ scalar reductions, independent of the constraint-matrix nonzero count and the number of source partitions. Increasing the number of sources or GPUs therefore increases only local computation while communication remains fixed. Although the dual update is serialized on rank 0, its cost is negligible relative to the parallel sparse matrix--vector products and projections.

\begin{table*}[t!]
\centering
\small
\begin{tabular}{@{}p{4.0cm}p{11.0cm}@{}}
\toprule
\textbf{Component} & \textbf{Interface / Semantics} \\
\midrule

\texttt{ObjectiveFunction} &
Encodes $(\mathbf{A}, \mathbf{b}, \mathbf{c})$.
\texttt{calculate($\boldsymbol{\lambda}$, $\gamma$)} returns
$(g(\boldsymbol{\lambda}), \nabla g(\boldsymbol{\lambda}), \mathbf{x}^{*}_{\gamma}(\boldsymbol{\lambda}))$,
where $\mathbf{x}^{*}_{\gamma}(\boldsymbol{\lambda}) =
\Pi_{\mathcal{C}}(-\frac{1}{\gamma}(\mathbf{A}^\top\boldsymbol{\lambda}+\mathbf{c}))$ and
$\nabla g = \mathbf{A}\mathbf{x}^{*}_{\gamma} - \mathbf{b}$.
Implements only tensor-level ops; reuses CSC layout (Sec.~\ref{sec:gpu-layout}),
fused projection (Sec.~\ref{sec:gpu-triton}), and block structure
(Def.~\ref{def:matching-A}). \\

\texttt{ProjectionMap} &
Blockwise projection operator $\Pi_{\mathcal{C}}$.
Provides batched primitives for simplex / box / box-cut (Sec.~\ref{sec:gpu-batching}).
User-defined only for new constraint families; execution and batching infrastructure are reused. \\

\texttt{Maximizer} &
Runs dual ascent on $\boldsymbol{\lambda}\ge 0$:
iterates \texttt{ObjectiveFunction.calculate}, applies accelerated update,
conditioning, and continuation (Sec.~\ref{sec:opt}),
and hides distributed execution (Sec.~\ref{sec:gpu-dist}). \\

\bottomrule
\end{tabular}
\caption{Programming model with shared ObjectiveFunction, projection, and optimization components.}
\label{tab:components}
\end{table*}

\section{Programming Model}
\label{sec:architecture}

The system is reusable across formulations only if new LPs can be expressed without modifying the solve loop or distributed runtime. We achieve this by exposing a boundary at the three primitives from Section~\ref{sec:dual}: sparse products $\mathbf{A}\mathbf{x}$ and $\mathbf{A}^{\top}\boldsymbol{\lambda}$, and the blockwise projection $\Pi_{\mathcal{C}}$. All algorithmic logic (dual ascent, gradient evaluation, column-sharded execution, and NCCL communication) is shared; only formulation-specific operators are implemented locally.

DuaLip-Scala instead places the boundary at the LP-family level: users select one of two declarative schemas, and gradient computation is dispatched via schema-specific implementations. This makes even small changes non-local. For example, adding a global-count constraint $\sum_{(i,j)\in E} x_{ij} \le M$ only augments $\mathbf{A}$ and $\boldsymbol{\lambda}$, leaving the dual algorithm unchanged, yet requires modifications to the schema, parser, gradient code, Spark execution plan, and dispatch logic. The abstraction therefore tracks problem families rather than computation. Our system lowers the boundary to the gradient computation itself, implemented as tensor operations over sparse and dense structures. New coupling constraints correspond to a local change: one additional dual coordinate, one term in $\mathbf{A}^{\top}\boldsymbol{\lambda}$, and one gradient contribution from $\mathbf{x}$. The solver, projections, and distributed execution remain unchanged, as they operate purely on $\mathbf{x}$ and $\boldsymbol{\lambda}$.

\section{Algorithmic Improvements}\label{sec:opt}
Our solver builds on the ridge-regularized dual ascent framework of Section~\ref{sec:preliminaries}, optimizing the smoothed dual $g(\lambdab)$ via first-order updates with gradient access $\nabla g(\lambdab)=\Ab\xb_\gamma^*(\lambdab)-\bb$. Prior Scala/Spark implementations required instance-specific tuning of AGD variants and regularization strength to obtain stable convergence. We instead adopt a single optimization strategy based on continuation over the ridge parameter $\gamma$.

The method solves a sequence of regularized problems, starting from a large $\gamma$ to improve conditioning and stabilize early iterations, and gradually decreasing it to the target objective. Each stage is warm-started from the previous dual iterate, ensuring continuity across the schedule. This removes the need for instance-specific optimizer tuning and improves robustness across heterogeneous matching workloads of Definition~\ref{def:matching-A}. In practice, it combines fast initial progress under strong regularization with accurate solutions at small $\gamma$.

Within each stage, we apply Jacobi preconditioning to improve conditioning by rescaling the complex constraints using the diagonal of the Hessian proxy induced by $\Ab$. This improves the effectiveness of first-order updates, particularly in poorly conditioned regimes at intermediate values of $\gamma$. Together, continuation and Jacobi preconditioning yield a stable optimization trajectory from heavily smoothed problems to the original objective, improving convergence without manual hyperparameter tuning. Further details are provided in Appendix~\ref{app:alg_enhance}.

\section{Experiments}

We evaluate our system along three axes: (i) system-level performance and multi-GPU scaling; (ii) comparisons with state-of-the-art GPU-solvers; and (iii) the effect of algorithmic enhancements, including preconditioning and regularization continuation. We further demonstrate numerical parity with the Scala-based solver in the Appendix Section~\ref{app:additional_experiments}. We use synthetic matching data to enable controlled scaling of problem size and sparsity. The data generation procedure and complete experimental setup are described in Section~\ref{app:experiment_setting} of the Appendix.

\subsection{System Performance and Scaling} 

\paragraph{Cross-Platform (CPU–GPU) Runtime and Scaling Efficiency} We compare per-iteration runtime between Scala (Spark-based) and the PyTorch-based GPU implementation. Using a fixed random seed, we generate identical problem instances compatible with the Scala solver input schema and evaluate both systems under equal configurations. Table~\ref{tab:pytorch_scala_time_comparison} reports the average time per algorithm iteration across 1000 iterations. For moderate problem sizes (25M sources), a single GPU already provides close to an order-of-magnitude improvement over Scala. As the problem size increases, model sharding across multiple GPUs becomes necessary to satisfy memory constraints. Beyond enabling larger instances, our multi-GPU parallelization significantly reduces iteration time, achieving more than 35$\times$ speedup relative to Scala and near-linear scaling within the GPU setting.

\begin{table}[t]
\centering
\sisetup{
    table-number-alignment = center,
    round-mode = places,
    round-precision = 2
}
\begin{tabular}{l S *{4}{S}}
\toprule
&  & \multicolumn{4}{c}{\textbf{DuaLip (PyTorch)}} \\
\cmidrule(lr){3-6}
\textbf{Sources} & \textbf{Scala}
& {\textbf{1 GPU}} 
& {\textbf{2 GPUs}} 
& {\textbf{3 GPUs}} 
& {\textbf{4 GPUs}} \\
\midrule
25M  & 2.46 & 0.27 & 0.14 & 0.09 & 0.07 \\
50M  & 3.44 & {-}    & 0.27 & 0.18 & 0.13 \\
75M  & 2.63 & {-}    & 0.43 & 0.29 & 0.21 \\
100M & 3.33 & {-}    & {-}    & 0.37 & 0.27 \\
\bottomrule
\end{tabular}
\caption{Average time per AGD iteration (seconds). Multi-GPU sharding enables larger instances and yields substantial speedups over the Spark-based Scala implementation.}
\label{tab:pytorch_scala_time_comparison}
\end{table}

\begin{figure}[t!]
      \centering
      \includegraphics[width=0.65\linewidth]{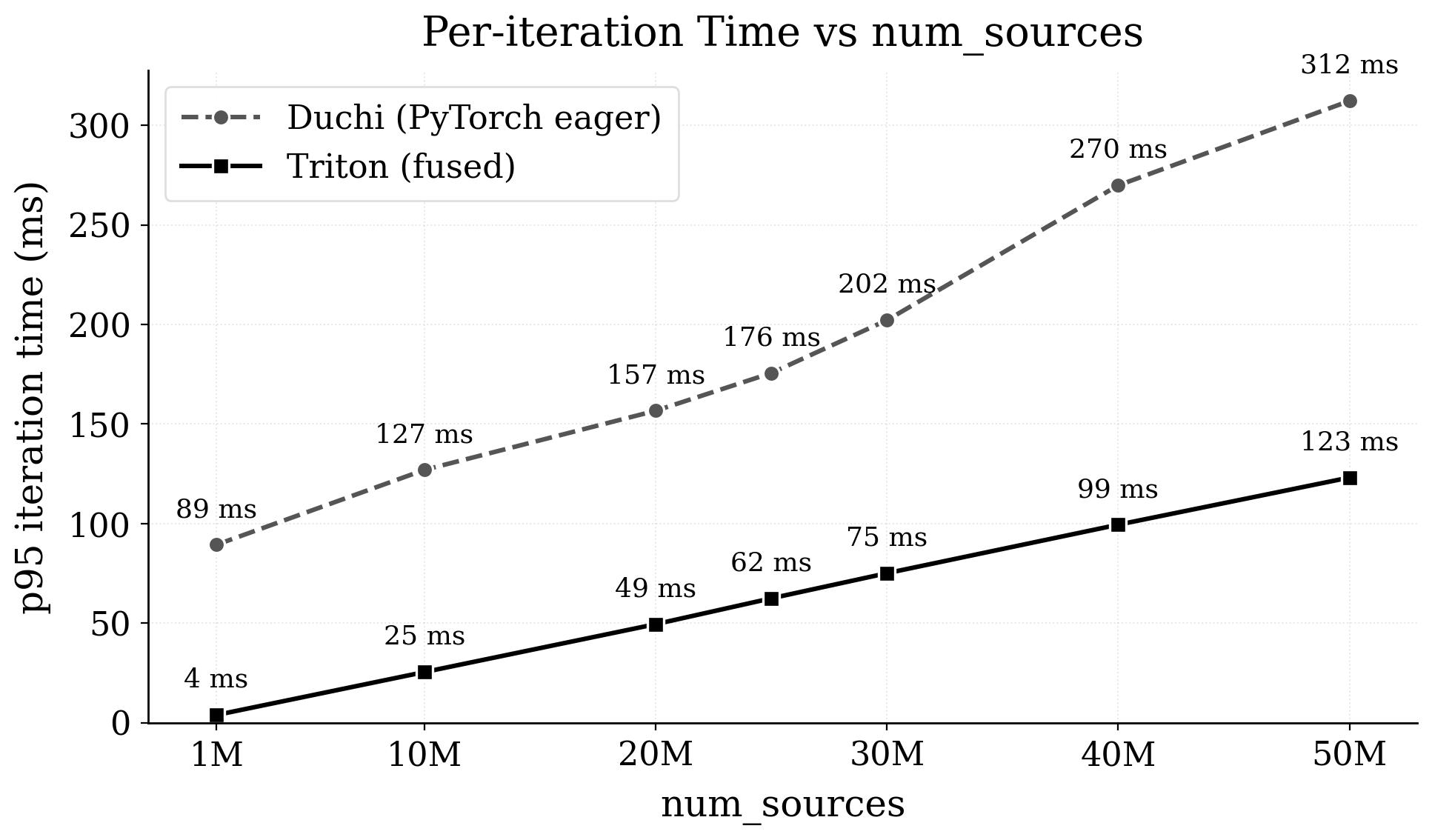}
      \\[4pt]
      \includegraphics[width=0.65\linewidth]{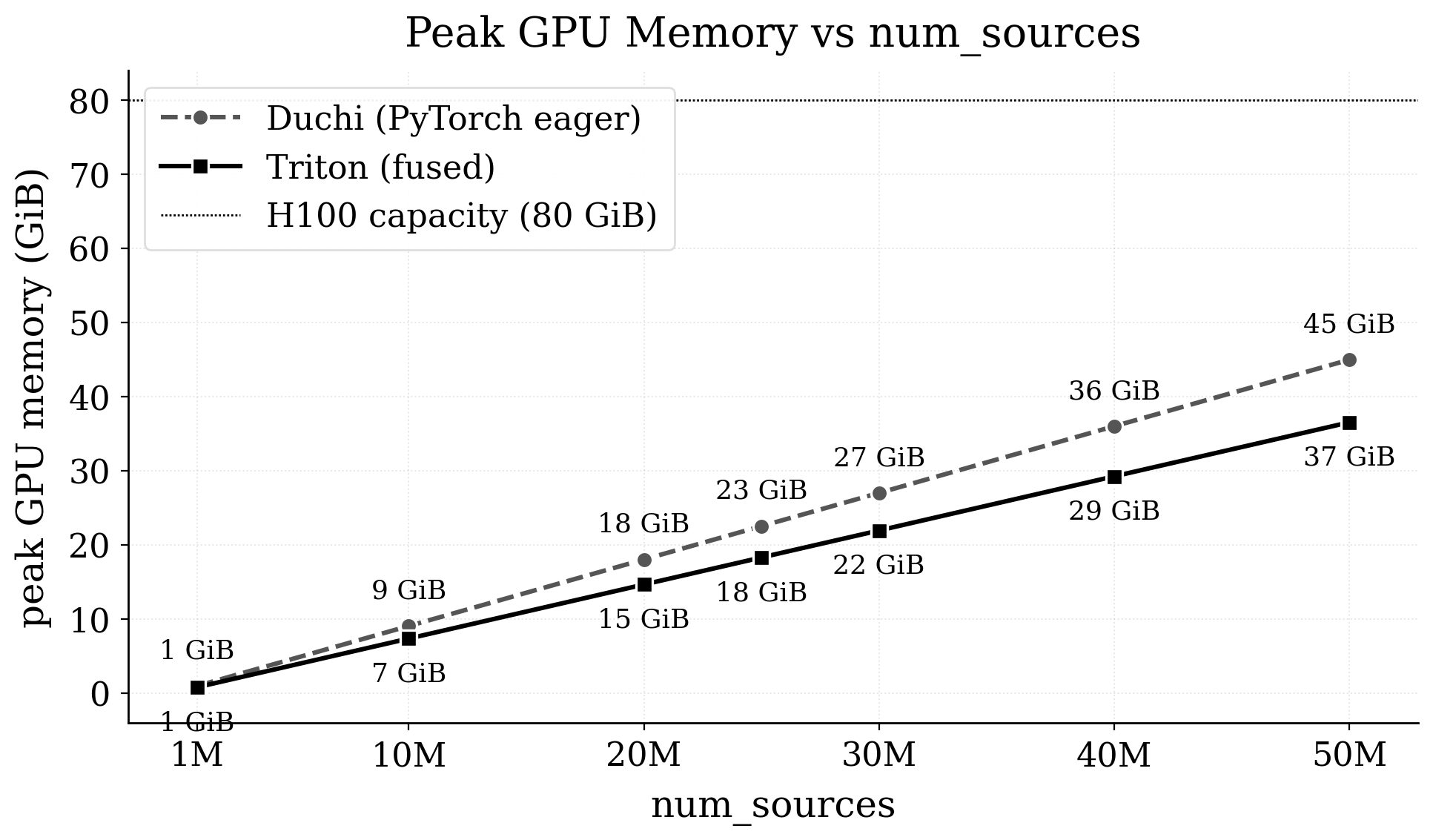}
      \caption{Triton fused projection vs.\ PyTorch-eager Duchi (single H100). (Top) p95 per-iteration time; (Bottom) peak GPU memory.}
      \label{fig:triton_kernel}
  \end{figure}

\paragraph{Kernel-Level Optimizations.}
We compare the fused projection kernel of Section~\ref{sec:gpu-triton}
against the multi-launch PyTorch reference implementation that it replaces.
Figure~\ref{fig:triton_kernel} reports per-iteration time and peak GPU memory on a single H100, sweeping problem sizes from $1$M to $50$M sources. The fused kernel achieves a $2.5\text{--}5\times$ per-iteration speedup for
$10$M–$50$M sources, and over $20\times$ on the $1$M instance, where kernel launch and dispatch overhead dominates the relatively small arithmetic
workload on per-column slices. Peak memory is reduced by a consistent
$\sim 20\%$ across all settings above $1$M, reflecting the removal of materialized intermediate tensors (sorted values, prefix sums, and masks)
from global memory. Both implementations scale linearly with problem size, so the memory reduction translates directly into an increase in the maximum single-GPU feasible instance before saturating the 80\,GiB device budget.

\begin{figure}[t!]
      \centering
      \includegraphics[width=0.65\linewidth]{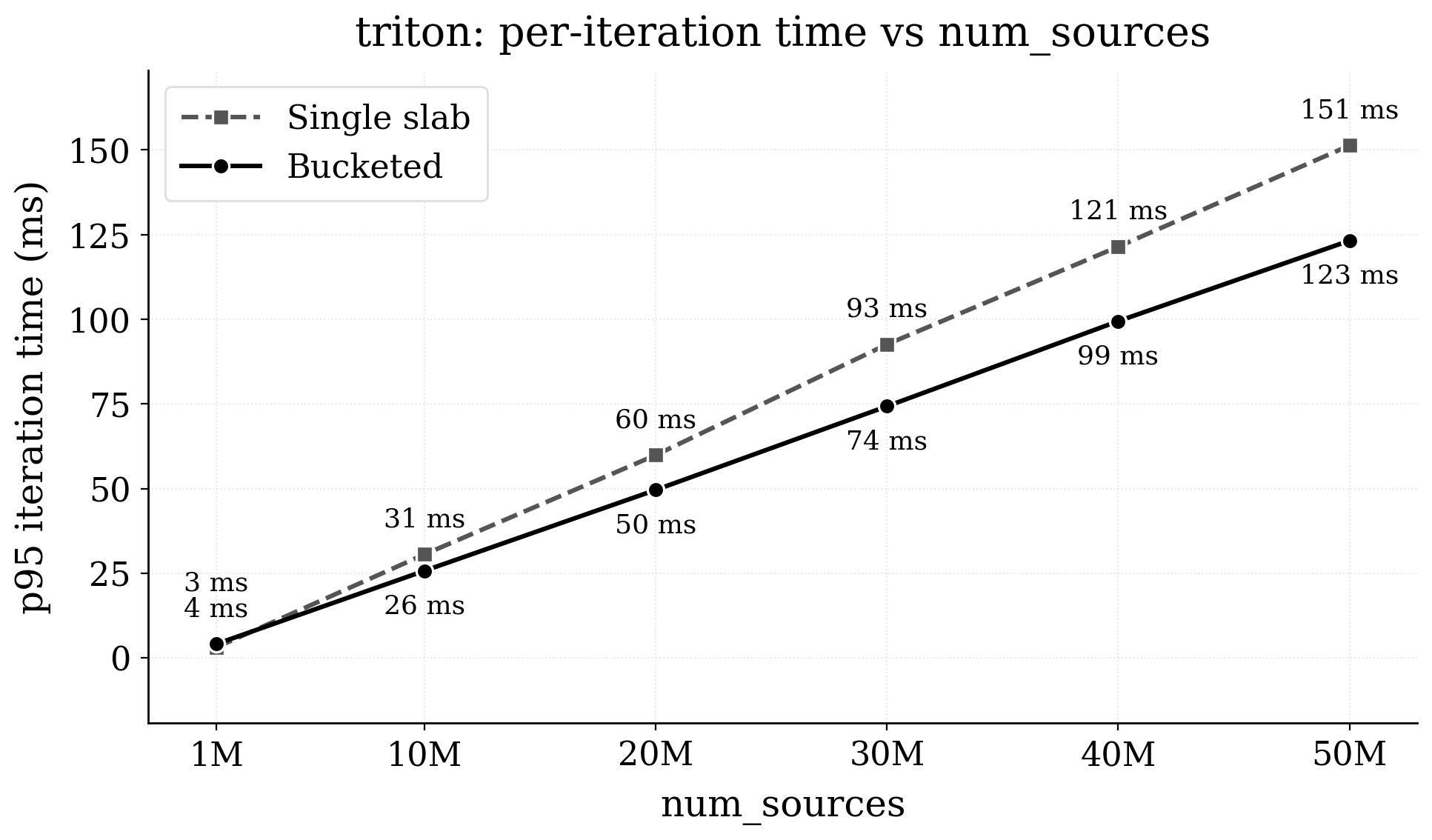}
      \\[4pt]
      \includegraphics[width=0.65\linewidth]{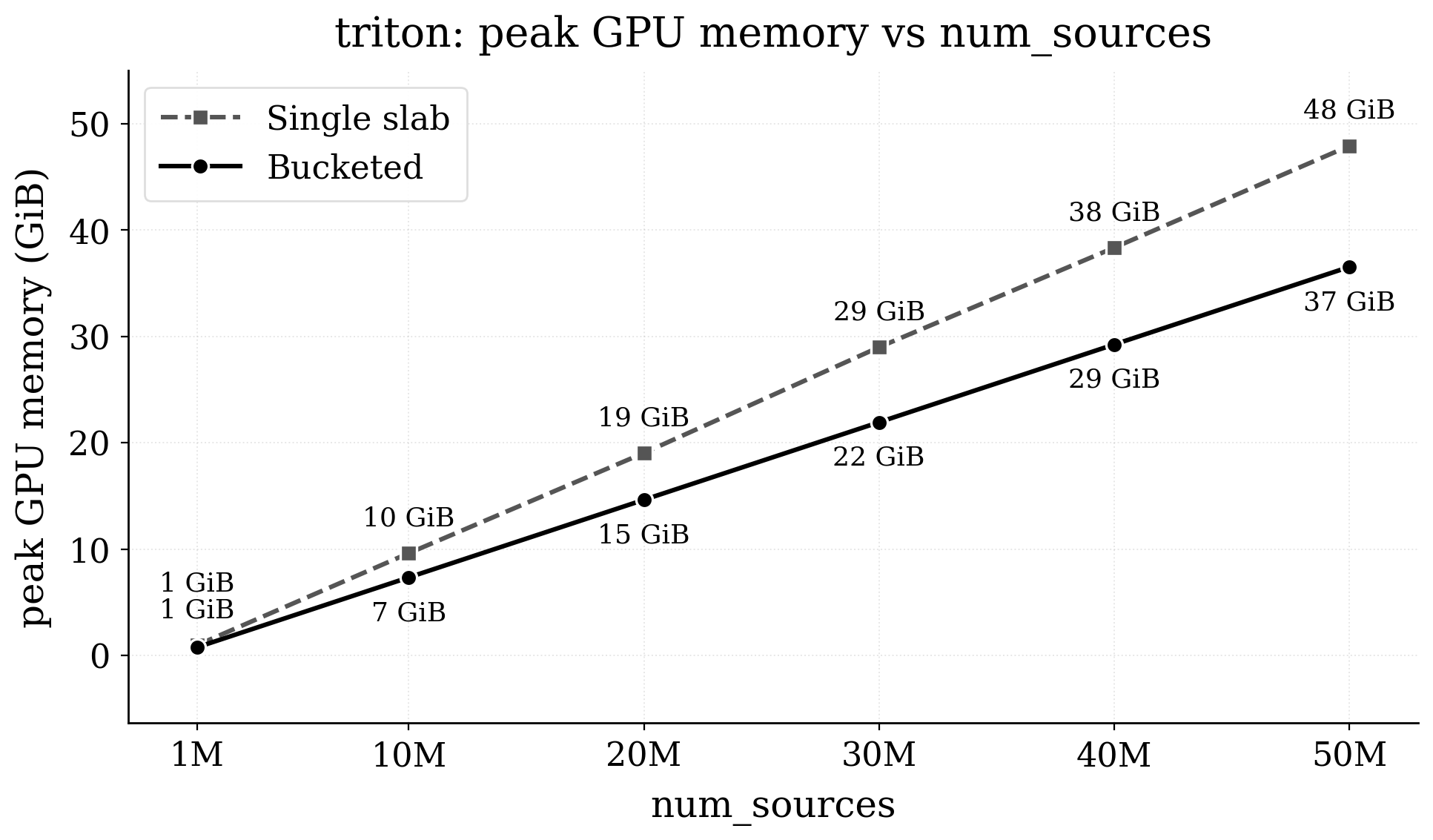}
      \caption{Geometric bucketing vs.\ single-slab baseline (Triton kernel, single H100). (Top) p95 per-iteration time; (Bottom) peak GPU memory.}
      \label{fig:batching}
      \vspace{-8pt}
  \end{figure}

\paragraph{Projection Batching.}
Figure~\ref{fig:batching} 
validates the
bucketing scheme described in Section~\ref{sec:gpu-batching}.
We compare the bucketed projection against the single-slab baseline
(\texttt{batching=False} in our implementation, corresponding to the
second failure mode in Section~\ref{sec:gpu-batching}) on top of the
Triton-fused kernel, sweeping problem size from $10$M to $50$M
sources. Bucketing delivers a consistent $\sim$$1.2\times$
per-iteration speedup by avoiding wasted compute on the zero-padded
entries of the single slab, and a consistent $\sim$$24\%$ reduction
in peak memory by replacing the global $[s_{\max} \times \mathrm{num\_sources}]$
slab with much smaller per-bucket slabs. Both gains are governed by
the skew of the per-source slice-length distribution: heavier tails
enlarge the fraction of arithmetic and memory the unbucketed path
spends on padding.

\begin{figure}[t]
      \centering
      \includegraphics[width=0.65\linewidth]{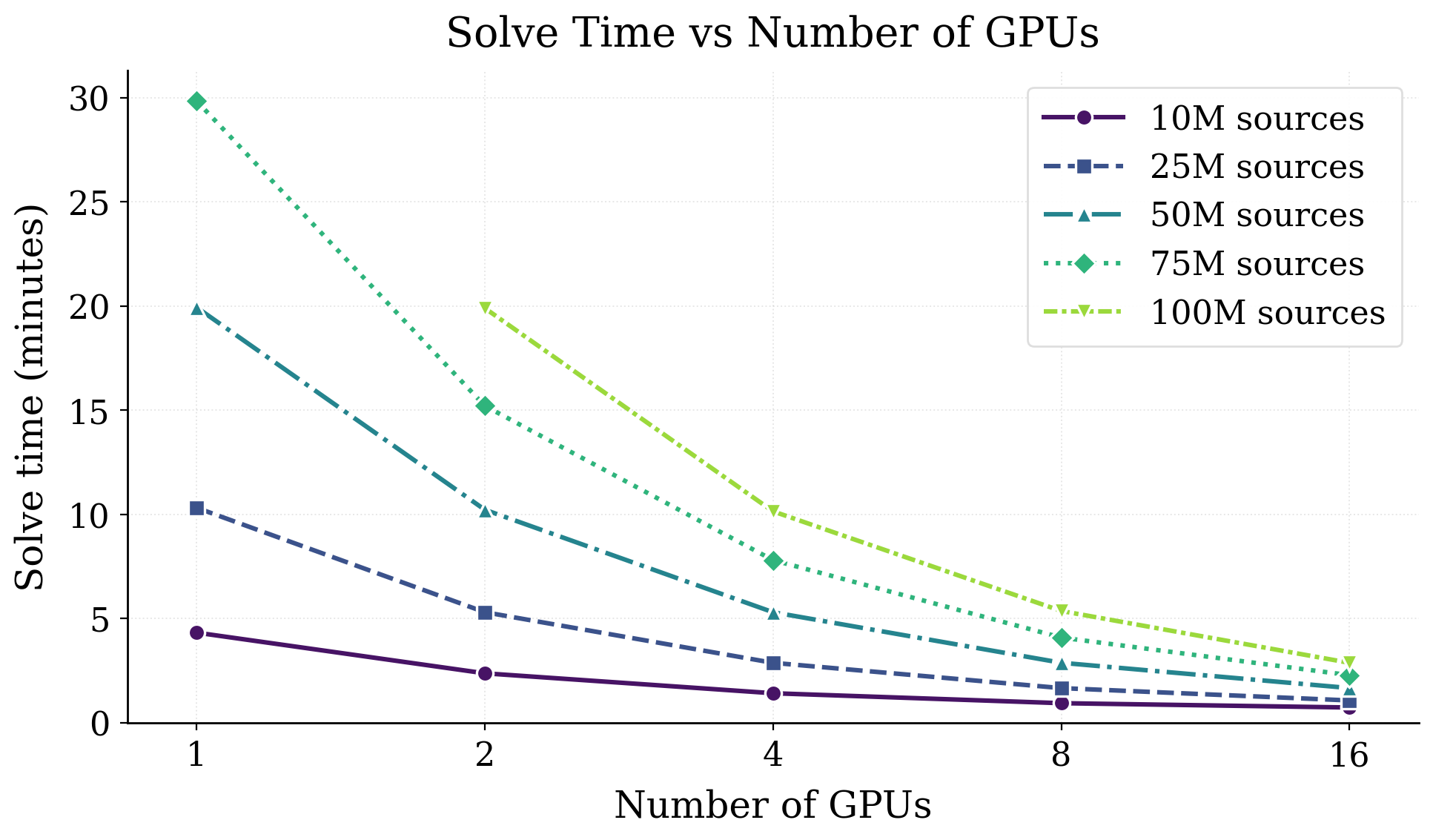}
      \\[4pt]
      \includegraphics[width=0.65\linewidth]{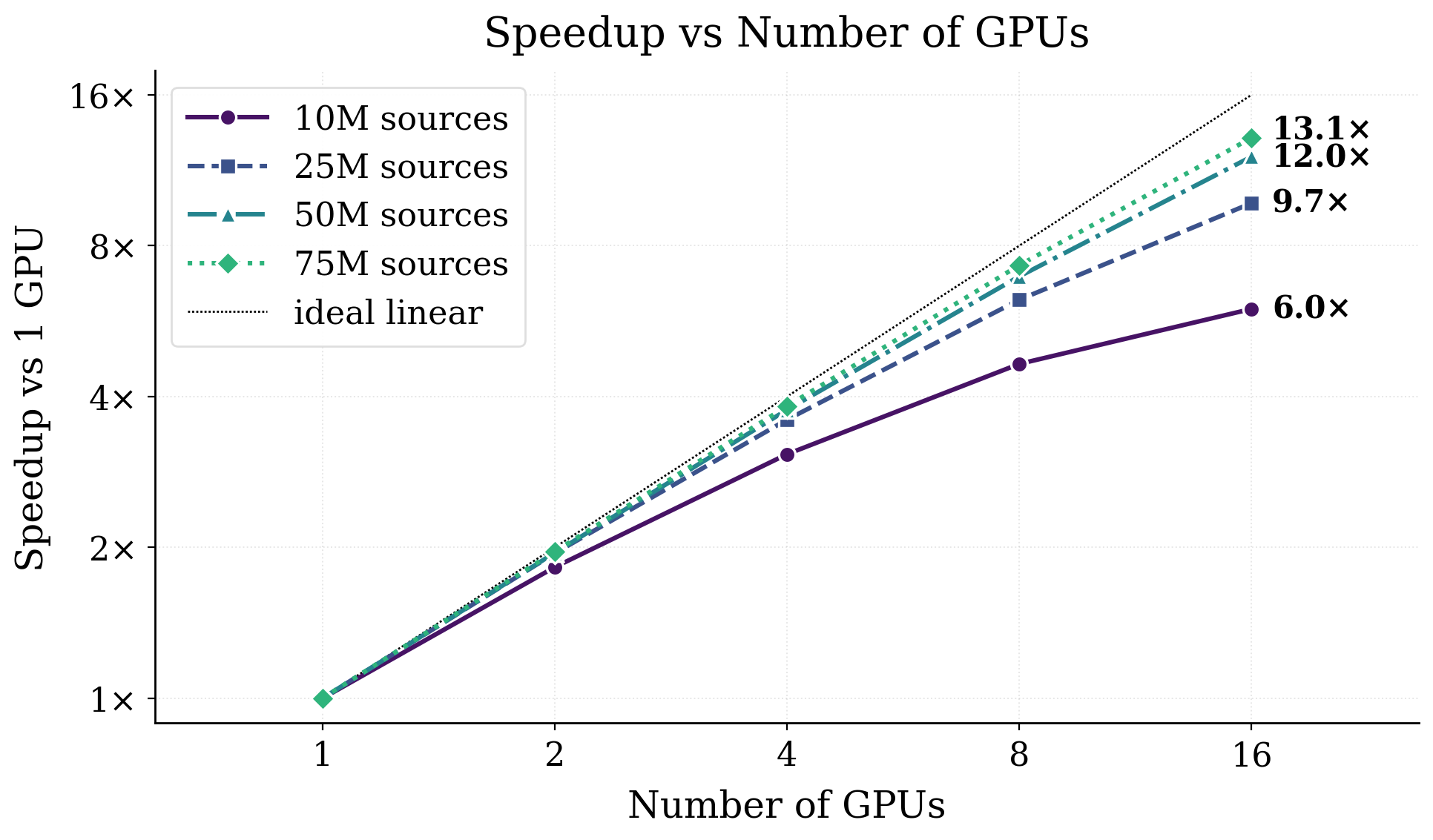}
      \caption{Scaling from 1–16 GPUs (16-GPU spans 2 H100 nodes). (Top) Solve time; (Bottom) speedup vs.\ 1 GPU.}
      \label{fig:scaling}
  \end{figure}

\paragraph{Multi-GPU and Multi-Node Scaling.}
We evaluate the distributed execution scheme of Section~\ref{sec:gpu-dist}
over $1$, $2$, $4$, $8$, and $16$ GPUs. The $1$–$8$ GPU configurations run
on a single H100 node, while $16$ GPUs span two nodes coordinated via a
\texttt{torchrun}-based launcher. All runs use the Triton-fused projection
with batching enabled. Figure~\ref{fig:scaling} reports end-to-end solve
time and speedup across problem sizes.

With a fixed destination count of $10{,}000$, per-iteration communication
is independent of source size (Section~\ref{sec:gpu-dist}), so scaling is
dominated by local gradient computation. At $16$ GPUs, we observe
$13.1\times$ speedup on $75$M sources ($82\%$ efficiency), $12.0\times$ on
$50$M ($75\%$), and $9.7\times$ on $25$M ($61\%$). Smaller problems saturate
earlier because the fixed reduce-and-broadcast overhead becomes a larger
fraction of each iteration ($6.0\times$ at $10$M, $37\%$ efficiency). The
scaling remains smooth across the $8 \rightarrow 16$ GPU transition,
indicating that the per-iteration communication—one
$|\boldsymbol{\lambda}|$-reduce and two $|\boldsymbol{\lambda}|$-broadcasts—
is well within NCCL bandwidth on the H100 interconnect. Beyond speedup, distributed execution enables instances that exceed single-GPU memory. The $100$M-source problem OOMs on a single device (working set exceeds 80\,GiB), but runs on $\geq 2$ GPUs due to linear column partitioning of the
constraint matrix. At $16$ GPUs, the $100$M-source instance completes in under $3$ minutes.

\begin{table*}[t!]
\centering
\caption{Solver runtime on synthetic statistical-matching LPs at scale.
Each cell is solve time in seconds on $N$ NVIDIA~H100~80GB GPUs (single node).
``OOM'' = the solver exhausted GPU/CPU memory and produced no output.
``--'' = configuration not run. Dualip generates the LP on the fly;
D-PDLP loads pre-built MPS files (135--327~GB on disk for the largest cases).}
\label{tab:runtime}
\setlength{\tabcolsep}{6pt}
\renewcommand{\arraystretch}{1.1}
\begin{tabular}{l r r S[table-format=4.0] S[table-format=4.0] S[table-format=4.1] S[table-format=4.1]}
\toprule
\multirow{2}{*}{Instance} & \multirow{2}{*}{$n \times m$} & \multirow{2}{*}{NNZ}
& \multicolumn{2}{c}{Dualip (s)} & \multicolumn{2}{c}{D-PDLP (s)} \\
\cmidrule(lr){4-5}\cmidrule(lr){6-7}
& & & {4 GPU} & {8 GPU} & {4 GPU} & {8 GPU} \\
\midrule
\multicolumn{7}{l}{\emph{Base instances}}\\
s50M\,d10K  &  $50\text{M}\times10\text{K}$  & $0.50\text{B}$ & 2384 & 1269 & 6.8  & 4.5 \\
s75M\,d10K  &  $75\text{M}\times10\text{K}$  & $0.75\text{B}$ & 3522 & 1841 & 10.2 & 6.6 \\
s80M\,d10K  &  $80\text{M}\times10\text{K}$  & $0.80\text{B}$ & 3738 & 1944 & 10.9 & 7.0 \\
s100M\,d10K & $100\text{M}\times10\text{K}$  & $1.00\text{B}$ & 4619 & 2441 & \multicolumn{1}{c}{\textbf{OOM}} & \multicolumn{1}{c}{\textbf{OOM}} \\
\midrule
\multicolumn{7}{l}{\emph{$\ell_1$-reformulated instances} ($\approx 2\times$ NNZ of base)}\\
s50M\,d10K + L1  &  $50\text{M}\times10\text{K}$  & ${\sim}1.0\text{B}$
& \multicolumn{1}{c}{NA} & \multicolumn{1}{c}{NA}
& \multicolumn{1}{c}{\textbf{OOM}} & \multicolumn{1}{c}{\textbf{OOM}} \\
s75M\,d10K + L1  &  $75\text{M}\times10\text{K}$  & ${\sim}1.5\text{B}$
& \multicolumn{1}{c}{NA} & \multicolumn{1}{c}{NA}
& \multicolumn{1}{c}{\textbf{OOM}} & \multicolumn{1}{c}{\textbf{OOM}} \\
s80M\,d10K + L1  &  $80\text{M}\times10\text{K}$  & ${\sim}1.6\text{B}$
& \multicolumn{1}{c}{NA} & \multicolumn{1}{c}{NA}
& \multicolumn{1}{c}{\textbf{OOM}} & \multicolumn{1}{c}{\textbf{OOM}} \\
s100M\,d10K + L1 & $100\text{M}\times10\text{K}$  & ${\sim}2.0\text{B}$
& \multicolumn{1}{c}{NA} & \multicolumn{1}{c}{NA}
& \multicolumn{1}{c}{\textbf{OOM}} & \multicolumn{1}{c}{\textbf{OOM}} \\
\bottomrule
\end{tabular}
\end{table*}

\begin{table*}[t!]
\centering
\caption{
Solution quality at 8 GPUs.
D-PDLP terminates when relative primal and dual residuals fall below $10^{-4}$.
Dualip runs 60,000 iterations with a six-stage $\gamma$ schedule
($\gamma \in \{10^{3}, 10^{2}, 10, 1, 10^{-1}, 10^{-2}\}$, 10,000 iterations per stage). All experiments completed within 2 hours time limit set for both solvers. ``Gap'' denotes relative objective gap.
Slack is the maximum constraint violation by each solver.
}
\label{tab:quality}
\setlength{\tabcolsep}{6pt}
\renewcommand{\arraystretch}{1.15}

\begin{tabular}{c l c c c c}
\toprule
Instance & Solver & Primal & Dual & Gap & Constraint Slack \\
\midrule

\multirow{2}{*}{s50M--d10K}
& D-PDLP  & $-2.510 \times 10^{6}$ & $-2.511 \times 10^{6}$
& $7.2 \times 10^{-5}$ & $4.6 \times 10^{-7\,\dagger}$ \\
& Dualip  & $-2.444 \times 10^{6}$ & $-2.444\times 10^{6}$
& $-6.9 \times 10^{-7}$ & $5.3 \times 10^{-2}$ \\
\addlinespace[3pt]

\multirow{2}{*}{s75M--d10K}
& D-PDLP  & $-3.829 \times 10^{6}$ & $-3.829 \times 10^{6}$
& $7.9 \times 10^{-5}$ & $4.4 \times 10^{-7\,\dagger}$ \\
& Dualip  & $-3.724 \times 10^{6}$ & $-3.724 \times 10^{6}$
& $3.5 \times 10^{-10}$ & $1.2 \times 10^{-5}$ \\

\addlinespace[3pt]

\multirow{2}{*}{s80M--d10K}
& D-PDLP  & $-4.095 \times 10^{6}$ & $-4.096 \times 10^{6}$
& $8.0 \times 10^{-5}$ & $4.1 \times 10^{-7\,\dagger}$ \\
& Dualip  & $-3.983 \times 10^{6}$ & $-3.983 \times 10^{6}$
& $5.6 \times 10^{-12}$ & $2.7 \times 10^{-6}$ \\

\addlinespace[3pt]

\multirow{2}{*}{s100M--d10K}
& D-PDLP & \multicolumn{4}{c}{\textbf{OOM}} \\
& Dualip & $-4.879 \times 10^{6}$ & $-5.012 \times 10^{6}$
& $1.6 \times 10^{-6}$ & $1.4 \times 10^{-1}$ \\

\bottomrule
\multicolumn{6}{l}{\footnotesize
$^{\dagger}$ D-PDLP reports a relative primal residual instead of max constraint slack.
}
\end{tabular}
\end{table*}

\subsection{Comparison with D-PDLP}
We compare our system against the state-of-the-art GPU primal--dual solver D-PDLP~\cite{li2026dpdlpscalingpdlpdistributed} on matching LPs generated with varying number of sources from $50$M to $100$M. Both solvers use float64 data type. The base formulation yields LPs with up to one billion nonzeros. We also consider an $\ell_1$-regularized variant of the problem, where the $\ell_2$-regularization in Problem~\ref{eq:primal} is replaced by a term $\gamma \|\mathbf{x}\|_1$ in the objective. This formulation can be equivalently expressed as a linear program by introducing auxiliary variables, and is useful for enforcing stability across consecutive solver runs.
All experiments run on a single node with NVIDIA H100\,80\,GB
GPUs (either $4\times$ or $8\times$ on the same host).  The solver is invoked
in single-node multi-GPU mode with AGD updates, a six-stage geometric
$\gamma$-schedule
$\gamma\in\{10^{3},10^{2},10,1,10^{-1},10^{-2}\}$ ($10{,}000$ iterations per
stage; $60{,}000$ total), Jacobi preconditioning, 
and AGD step-size range $[10^{-5},10^{-1}]$. D-PDLP is run on the same instance with presolve disabled and a
$10^{-4}$ relative tolerance on the optimality and feasibility residuals. Presolve is disabled to allow more memory headroom for D-PDLP solver. 
Dualip-DPU materializes the LP directly on the GPU; D-PDLP first loads a
pre-generated MPS file the size of which grows from
63\,GB (s50M base) to 306\,GB (s100M $\ell_1$-reformulated).

Table~\ref{tab:runtime} reports end-to-end solve time on the eight problem
instances and two GPU counts.
Two observations stand out.
\emph{First,} Dualip completes \emph{every} base instance up to $10^{9}$
nonzeros at both 4 and 8 GPUs, while D-PDLP runs out of memory on the
largest base instance (s100M, $1.0\times\!10^{9}$ nonzeros) and on
\emph{every} $\ell_1$-reformulated instance, including the smallest
($\sim\!10^{9}$ nonzeros). 
\emph{Second,} the runtime growth on the instances Dualip solves is
near-linear in problem size and in hardware: doubling the nonzero count from $0.50\text{B}$ to $1.00\text{B}$ increases 8-GPU solve
time by $1.9\times$, and doubling the GPU count from $4$ to $8$ yields a
$1.88$--$1.92\times$ speedup across the four base sizes.  Where D-PDLP
succeeds it is markedly faster per instance, but this efficiency does not
translate into a usable solver at the scales required by the $\ell_1$ reformulation, which is the operating point of interest for our application. Note that the GPU-based DuaLip solver can solve such instances within a 2-hour time budget, making it practical for production use cases where this level of latency is acceptable.

Table~\ref{tab:quality} compares the solutions returned by the two solvers
at 8 GPUs on the instances where D-PDLP completes, together with the
largest instance (s100M, $1.0\!\times\!10^{9}$ nonzeros) which only Dualip
solves.  D-PDLP terminates adaptively once both the relative primal and
dual residuals fall below $10^{-4}$.  Dualip runs the full $60{,}000$-iteration schedule described
above; we report the primal and dual objectives
at the end of the schedule, the absolute primal--dual gap, and the maximum
positive constraint violation (``slack'') of the recovered primal. On instances where both solvers succeed, the dual objectives agree to four significant figures ($\Delta \mathrm{dual}/|\mathrm{dual}| < 10^{-6}$ across all three problem sizes), indicating that both methods converge to the same optimum when the regularization parameter is sufficiently small (0.01 in the final stage). This result is important because it shows that the regularized formulation provides a controllable trade-off: the regularization strength can be tuned to adjust solution fidelity while also improving numerical stability when needed.
Additionally, our system achieves a substantially smaller primal–dual gap (down to $10^{-12}$) compared to D-PDLP, reflecting the improved conditioning of the smoothed problem formulation. 

\subsection{Effect of Algorithmic Enhancements}

We now isolate the impact of two algorithmic improvements introduced in Section~\ref{sec:opt}: diagonal preconditioning and regularization continuation.

\begin{figure}[ht!]
    \centering
    \includegraphics[width=0.65\linewidth]{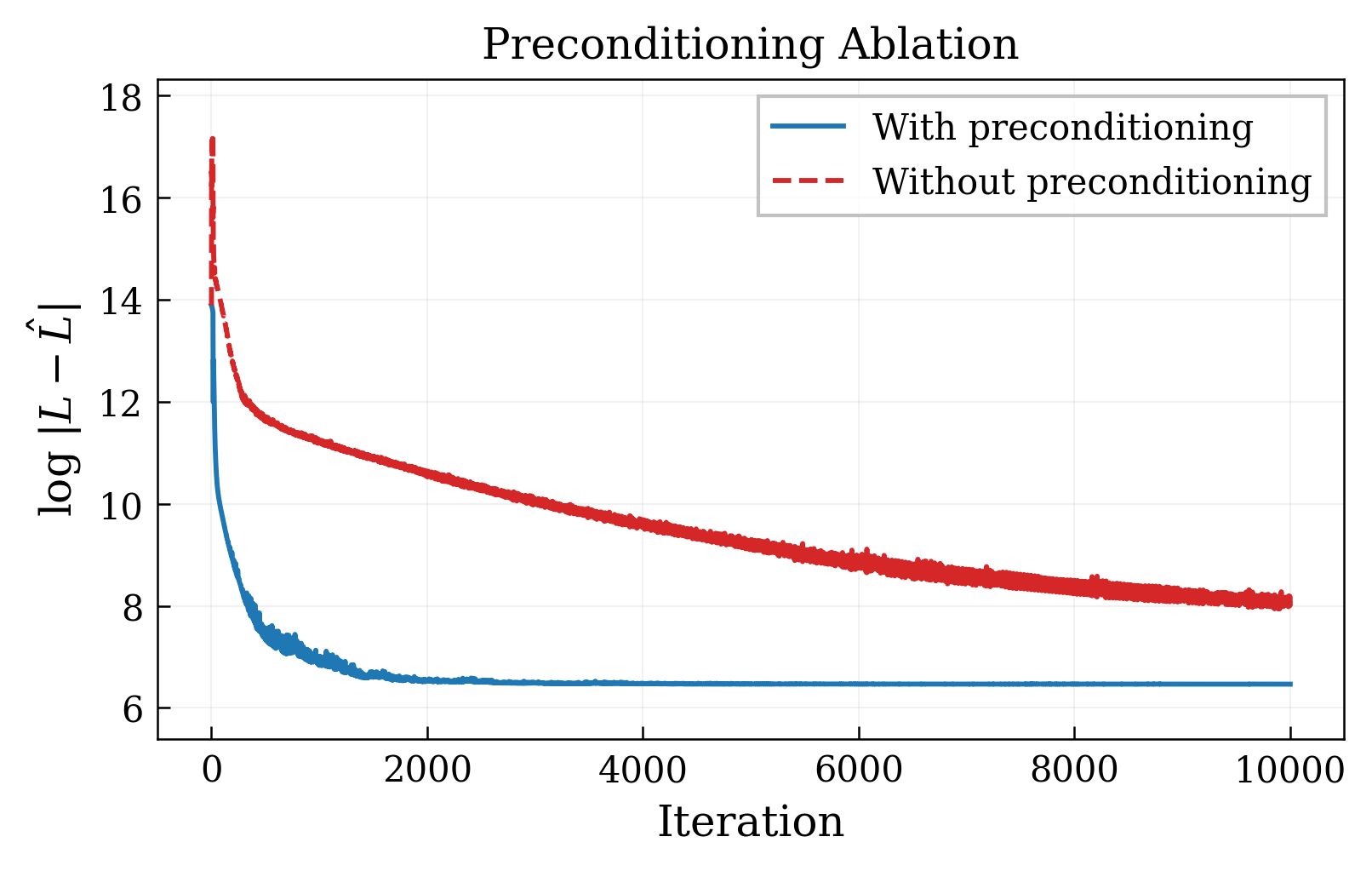}
    \caption{Effect of diagonal preconditioning. We report $\log(|L - \hat{L}|)$ for a 25M-source instance (10k destinations, 0.1\% sparsity). Preconditioning significantly improves early-stage convergence.}
    \label{fig:preconditioning}
\end{figure}

\paragraph{Preconditioning.}
Figure~\ref{fig:preconditioning} plots $\log(|L - \hat{L}|)$, where $L$ is the dual objective and $\hat{L}$ is the converged reference value. With diagonal preconditioning, convergence accelerates substantially, particularly in early iterations. This confirms that scaling the dual variables mitigates ill-conditioning arising from heterogeneous constraint magnitudes.

\begin{figure}[ht!]
    \centering
    \includegraphics[width=0.65\linewidth]{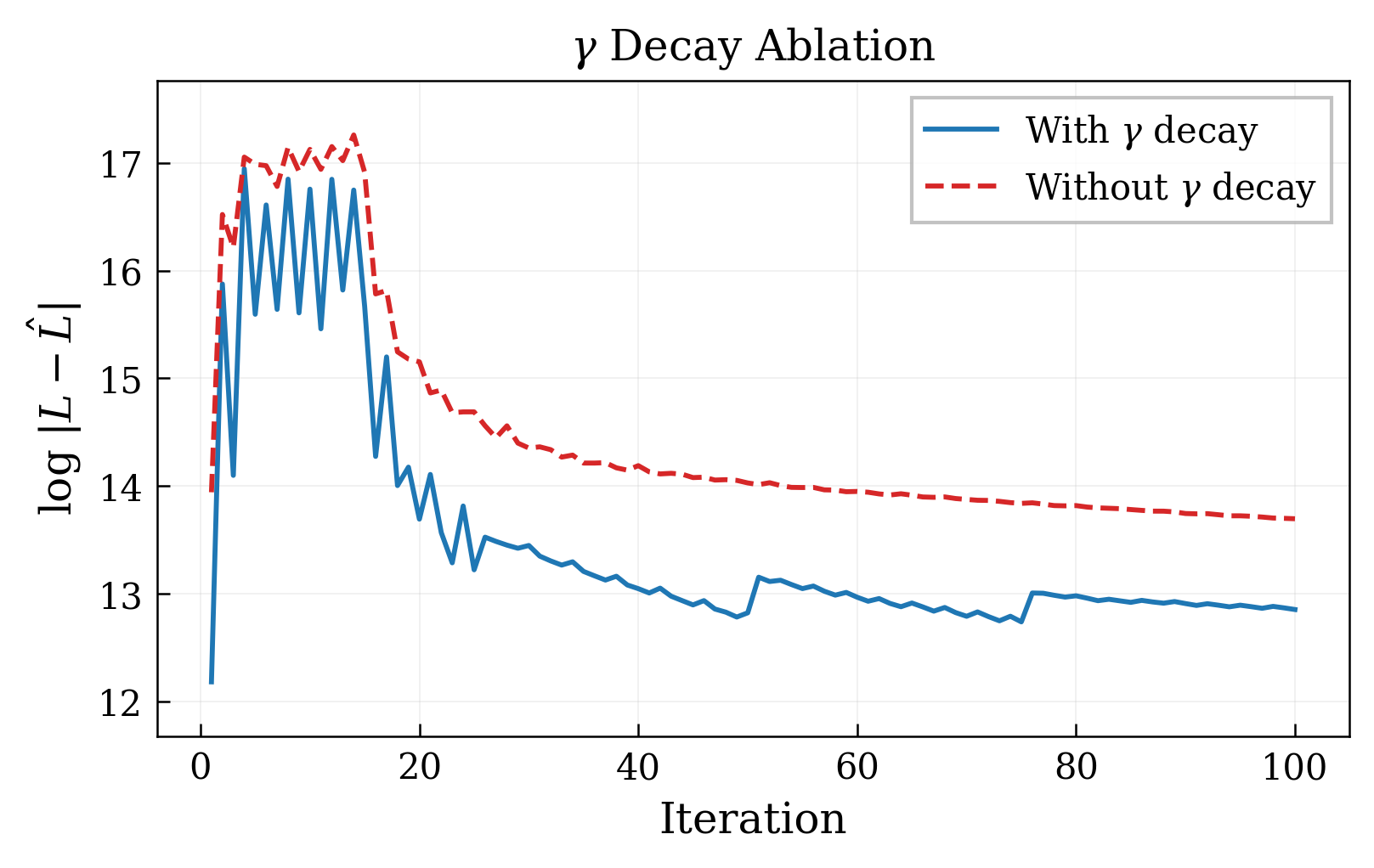}
    \caption{Effect of regularization continuation. Decaying $\gamma$ during optimization accelerates convergence while preserving solution fidelity.}
    \label{fig:gamma_decay}
\end{figure}

\paragraph{Regularization Continuation.}
Figure~\ref{fig:gamma_decay} evaluates the continuation strategy for the ridge parameter $\gamma$. Starting from a larger $\gamma$ stabilizes and accelerates early optimization, while gradual decay ensures the final solution closely approximates the unregularized LP optimum. Decaying $\gamma$ from 0.16 to 0.01 (halved every 25 iterations) yields faster convergence compared to using a fixed regularization level.

\section{Discussion and Limitations}

In this paper, we described the main architectural, algorithmic, and systems design choices for extreme-scale matching LPs in production systems. A natural next step is to evaluate the same dual-ascent framework on broader classes of linear programming benchmarks beyond matching. A key limitation is the lack of publicly available datasets at the scale and structure of production matching workloads. As a result, evaluation is based primarily on large synthetic instances calibrated to observed industrial distributions. Developing standardized benchmark suites for extreme-scale matching LPs would significantly improve reproducibility and enable more systematic comparison of methods in this regime.

\newpage
\appendix
\section{Experiment setting}\label{app:experiment_setting}

\paragraph{Synthetic LP construction.}
We construct synthetic instances by first generating a sparse bipartite interaction graph and then assigning values and constraint coefficients on its edges. Given a target number of ``requests'' \(I\), ``resources'' \(J\), and a target sparsity level, we draw a lognormal ``breadth'' parameter for each resource \(j\), normalize these to obtain probabilities \(p_j\), and sample the number of incident requests \(K_j \sim \mathrm{Poisson}(p_j I \nu)\), truncated at \(I\), where $\nu$ is the desired average number of nonzeros per row. For each resource \(j\), we then select \(K_j\) distinct requests and create edges \((i,j)\). On each edge, we draw a resource-specific value scale \(v_j\), a request-specific responsiveness factor \(u_i\), and multiplicative noise \(\varepsilon_{ij}\), and define a nonnegative value coefficient
\[
c_{ij} = \min\bigl(v_j u_i \varepsilon_{ij},\, c_{\max}\bigr).
\]
Constraint coefficients are taken to be scaled versions of these values,
\[
a_{ij} = s_j c_{ij},
\]
where the per-resource scale \(s_j\) is also drawn from a lognormal distribution. This construction yields a sparse matrix \(\Ab\) whose rows differ both in support size and magnitude (often by several orders), and a matching value matrix \(\Cb\) with the same sparsity pattern. 

\paragraph{Source capacities and right-hand side.}
Right-hand side source capacities \(b_j\) are chosen to make a nontrivial fraction of constraints active. Instead of taking \(b_j\) proportional to the sum \(\sum_i a_{ij}\), we approximate the maximum feasible load each resource could receive under the per-request simplex constraint (each request can allocate to at most one resource) by a greedy assignment: for each request \(i\), we identify the incident edge with the largest \(a_{ij}\) and assign that amount to the corresponding resource. Summing these contributions over requests gives a ``greedy load'' \(\ell_j\) for each resource, and we set 
\[
b_j = \rho_j \bigl(\ell_j + \varepsilon\bigr),
\]
where \(\rho_j\) is drawn uniformly from \([0.5, 1.0]\) and \(\varepsilon > 0\) is a small constant. This ensures that some resource constraints are binding while others remain slack in the optimal solution. The final LP data passed to the solver is the sparse CSC representation of $\Ab$, the corresponding value matrix (with signs adjusted to match our minimization convention), and the source capacity vector $\bb$.

\section{Additional Experiments}\label{app:additional_experiments}

\subsection{Implementation parity.}
We first verify numerical equivalence between the PyTorch implementation and the original Scala solver under identical problem instances and hyperparameters. Both systems are initialized from the same primal-dual state and execute the same accelerated gradient descent (AGD) update schedule, with differences limited to execution backend (CPU vs. GPU) and memory layout.

\begin{figure}[ht!]
    \centering
    \includegraphics[width=0.95\linewidth]{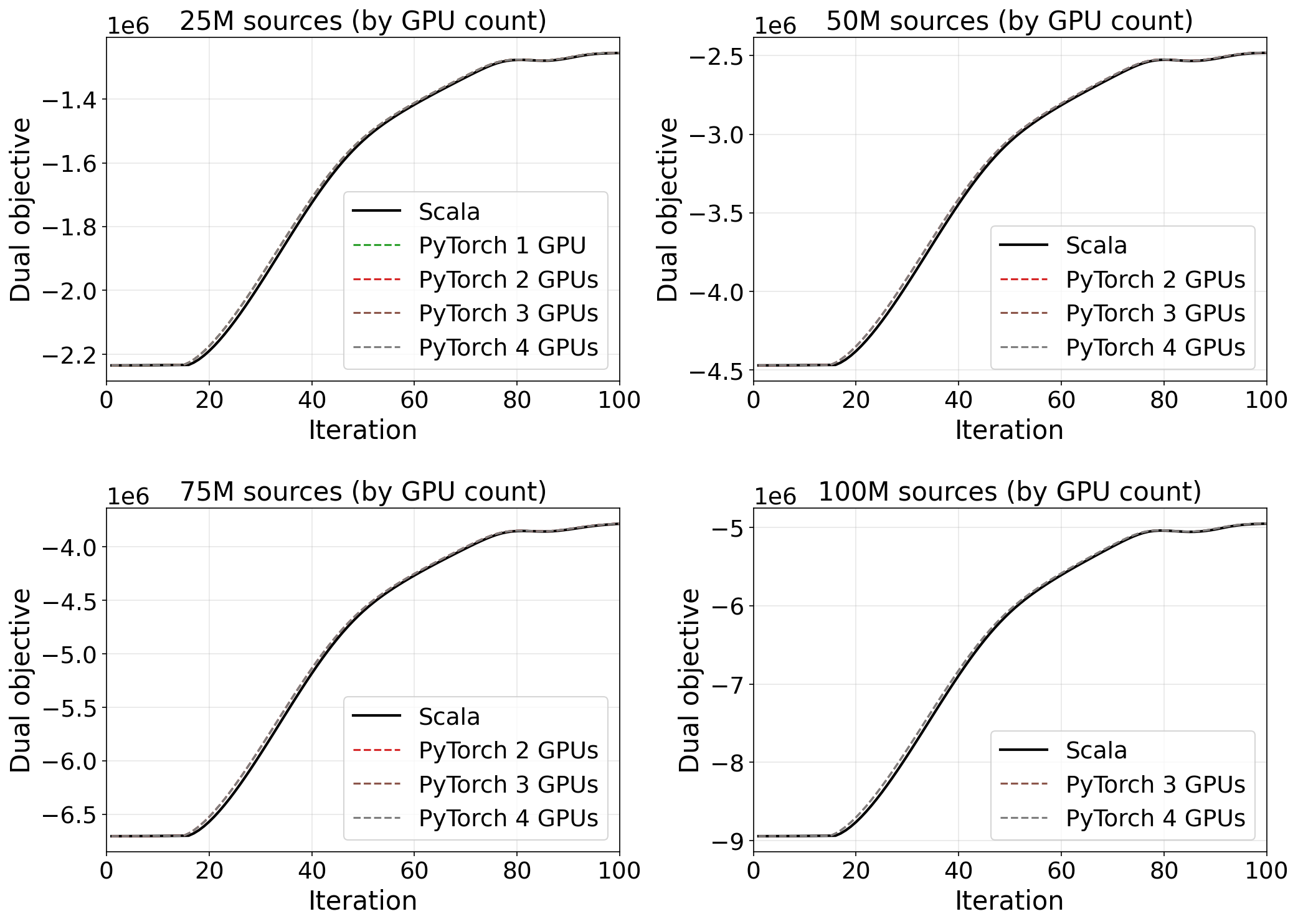}
    \caption{Scala–DuaLip (PyTorch) parity. Each panel shows the dual objective versus AGD iteration for the Scala and PyTorch implementations. The near-perfect overlap confirms numerical equivalence.}
    \label{fig:scala_pytorch_parity}
\end{figure}

\begin{figure}[ht!]
    \centering
    \includegraphics[width=0.95\linewidth]{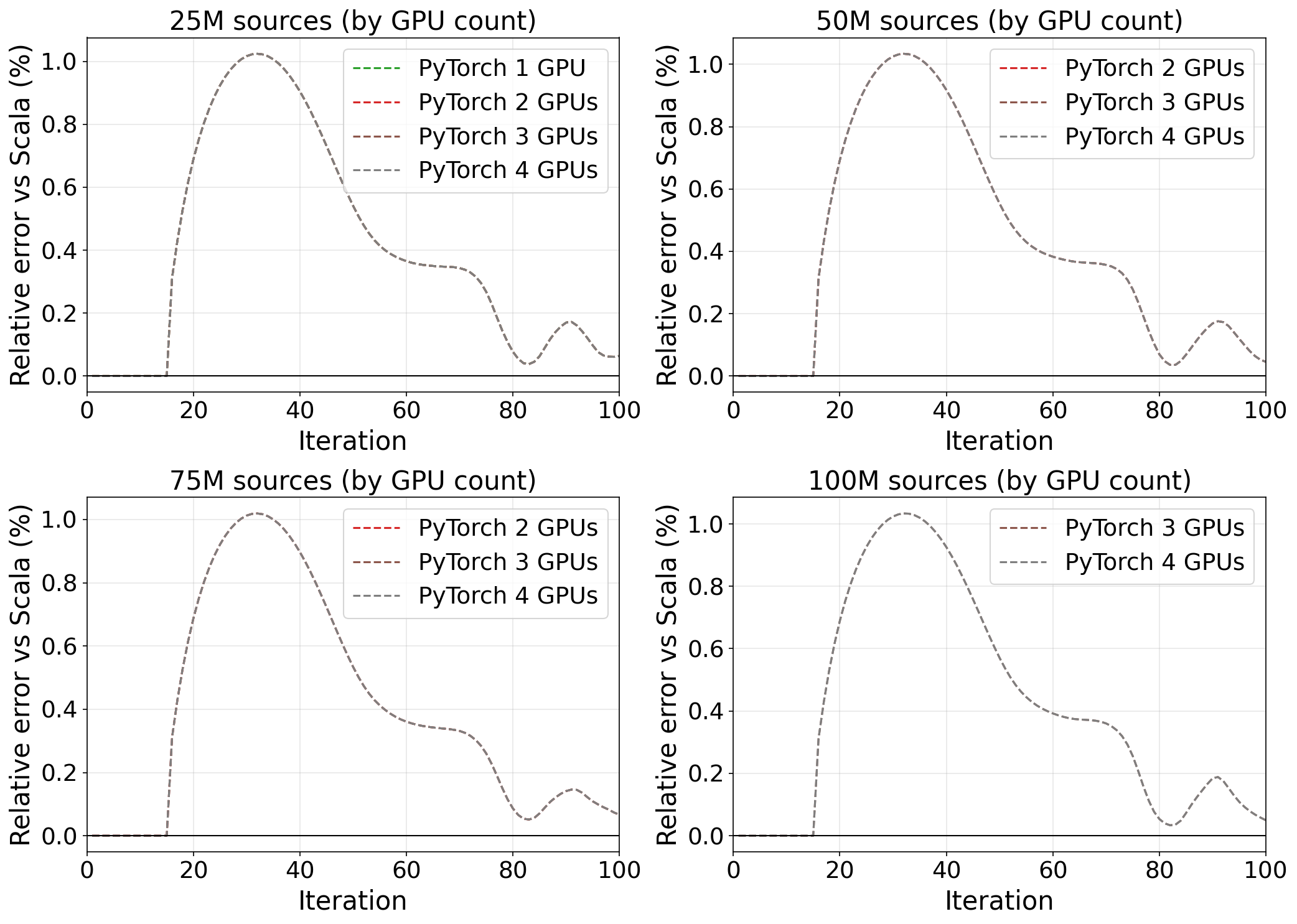}
    \caption{Relative error in dual objective compared to the Scala solver. The error drops below 1\% within the first 100 iterations across all settings.}
    \label{fig:scala_pytorch_relative_error}
\end{figure}

Figure~\ref{fig:scala_pytorch_parity} compares dual objective trajectories across iterations in single-GPU and multi-GPU configurations. Across all tested problem sizes and constraint families, the trajectories closely overlap, indicating that distributed execution and kernel fusion do not alter the optimization dynamics. To quantify discrepancies, Figure~\ref{fig:scala_pytorch_relative_error} reports the relative error in the dual objective with respect to the Scala solver, computed as
\[
\frac{|g_{\text{torch}}(\lambda_t) - g_{\text{scala}}(\lambda_t)|}{|g_{\text{scala}}(\lambda_t)|}.
\]
In all configurations, the relative error drops below 1\% within the first 100 iterations and continues to decay as the iterates converge, consistent with stabilization of the primal-dual dynamics.

We further observe that error behavior is stable across single-GPU and multi-GPU runs, indicating that NCCL-based aggregation and column sharding preserve numerical consistency. These results confirm that the PyTorch implementation faithfully reproduces the optimization trajectory of the production Scala system while supporting distributed execution and GPU acceleration.

\subsection{Algorithm enhancements}\label{app:alg_enhance}

A central area of focus in improving the DuaLip solver was to improve the performance and robustness of the ridge-regularized dual ascent method that underpins this line of work. The basic framework of Section~\ref{sec:preliminaries} admits many first-order instantiations; in practice, the dominant convergence issues we observed on production matching workloads stemmed from three sources: (i) poor conditioning of $\Ab \Ab^\top$ in the dual, (ii) tradeoff between convergence speed and solution quality with the ridge regularization term and (iii) heterogeneous scales in the primal variables that interact badly with the quadratic regularizer. We address these in turn via Jacobi-style row normalization and a regularization schedule. 


\paragraph{Jacobi preconditioning / row normalization.}
Our first enhancement is to improve the conditioning of the dual problem by rescaling the complex constraints. We assume $\Ab$ is full row rank. Intuitively, when some rows of $\Ab$ have much larger norms than others, gradient steps for the smoothed dual move too cautiously in some directions and too aggressively in others. 

When the projection operator is inactive (or the identity), the dual gradient reduces to
\[
\nabla \gb(\lambdab)\;=\;-\frac{1}{\gamma}\bigl(\Ab\Ab^\top\lambdab+\Ab\cbb\bigr)\;-\;\bb,
\]
so the Hessian is
\[
\nabla^2 \gb(\lambdab)\;=\;-\frac{1}{\gamma}\,\Ab\Ab^\top.
\]
The convergence of first-order methods on $g$ is therefore governed by the conditioning of $\Ab\Ab^\top$.

We apply a standard row-scaling transform. Let
\[
\Db=\mathrm{diag}\!\bigl(\|\Ab_{1*}\|_2^{-1},\ldots,\|\Ab_{m*}\|_2^{-1}\bigr)
\]
for all rows with nonzero norm (rows with $\|\Ab_{r*}\|_2=0$ are redundant and may be dropped or left unscaled with $\Db_{rr}=1$), and define the row-scaled system
\[
\Ab'=\Db\Ab,\qquad \bb'=\Db\bb.
\]
Because $\Db$ has positive diagonal entries, row scaling preserves the feasible set exactly:
\[
\{\,\xb:\Ab\xb\le\bb\,\}\;=\;\{\,\xb:\Ab'\xb\le\bb'\,\}.
\]
Moreover,
\[
\Ab'\Ab'^\top\;=\;\Db(\Ab\Ab^\top)\Db,
\qquad
\Db^2=\mathrm{diag}(\Ab\Ab^\top)^{-1}\ \ \text{(on nonzero rows)},
\]
so row normalization is precisely Jacobi preconditioning of the dual Hessian $-\nabla^2 \gb(\lambdab)=\frac{1}{\gamma}\Ab\Ab^\top$.

For the matching constraint matrix of Definition~\ref{def:matching-A}, $\Ab$ is a horizontal concatenation of diagonal subblocks across sources, so $\Ab\Ab^\top$ is a sum of (nearly) diagonal matrices (one per source) and is therefore close to diagonal in practice. Enforcing $\mathrm{diag}(\Ab'\Ab'^\top)=\Ib$ tightly clusters the spectrum; in the ideal diagonal/orthogonal case it yields $\Ab'\Ab'^\top=\Ib$ and condition number~$1$.

The following lemma formalizes this intuition under a simple statistical model for the matching blocks.

\vspace{0.5cm}
\begin{lemma}
\label{lemma:precon}
Let $\Ab=[\Ab_1~\cdots~\Ab_I]\in\R^{mJ\times IJ}$ with user blocks
$\Ab_i\in\R^{mJ\times J}$ that are i.i.d.\ across $i$ and diagonal by rows as in
Definition~\ref{def:matching-A}. Let $\Db_{\mathrm{exp}}=\mathrm{diag}\!\big(\EE\|\Ab_{1*}\|_2^2,\ldots,\EE\|\Ab_{m_2*}\|_2^2\big)^{-1/2}$
and $\widetilde{\Ab}=\Db_{\mathrm{exp}}\Ab$. Then $\mathrm{diag}\!\big(\EE[\widetilde{\Ab}\widetilde{\Ab}^\top]\big)=\Ib$.
If, in addition, for $r\neq s$,
\[
\EE\!\big[\langle \Ab_{r*},\Ab_{s*}\rangle\big]
\;\le\; \eta\,\sqrt{\EE\|\Ab_{r*}\|_2^2\,\EE\|\Ab_{s*}\|_2^2}
\quad\text{with }\eta\in[0,1),
\]
then
\[
\kappa\!\big(\EE[\widetilde{\Ab}\widetilde{\Ab}^\top]\big)
\;\le\; \frac{1+(m-1)\eta}{\,1-(m-1)\eta\,}.
\]
\end{lemma}

Thus, under mild cross-row correlation assumptions, row normalization nearly equalizes the eigenvalues of $\Ab\Ab^\top$ in expectation for matching workloads, which stabilizes dual first-order updates.

\paragraph{Regularization decay} The preconditioning above addresses the dominant factor when the simple constraints are inactive: in that regime, the condition number of $\Ab\Ab^\top$ controls the convergence rate of first-order methods. However, the ridge parameter $\gamma$ enters separately through the smoothed dual. 

Next, we consider how to set the regularization hyperparameter over the course of optimization. While $\gamma$ does not affect the conditioning of $\Ab\Ab^\top$, it has a substantial effect on the smoothness of the objective. On the one hand, one prefers small $\gamma$ to avoid perturbing the original LP. Lemma 2 in \citet{basu2020eclipse} guarantees that there is always a range of sufficiently small $\gamma$ values that allow exact recovery of a solution to the un-smoothed LP. On the other hand, the Lipschitz constant of the gradient is only bounded by $\|\Ab\|_2^2/\gamma$ (see Lemma 3 in \cite{basu2020eclipse}), so very small $\gamma$ leads to a poorly conditioned dual problem and slow progress in practice.

To balance convergence speed and solution fidelity, we implement a simple continuation scheme in which $\gamma$ is initialized at a moderately large value (for stable, fast early progress) and then decayed on a pre-specified schedule and rate as the algorithm approaches a good dual solution. Since $\gamma$ directly affects the smoothness of the dual objective, we scale the maximum AGD step size proportionally with the decay of $\gamma$ to maintain stability across transition points. Overall, this approach allows for faster convergence in early iterations, while ensuring the final solution is not greatly perturbed from the true LP solution.

\newpage
\bibliographystyle{ACM-Reference-Format}
\bibliography{paper}
\end{document}